\documentclass[twoside,12pt]{article}
\usepackage{epsf,epsfig,amssymb,amsmath,graphicx,float}
\usepackage{color}
\usepackage{hyperref}
\usepackage{indentfirst}
\hypersetup{colorlinks,citecolor=blue,linkcolor=red,urlcolor=red}

\usepackage{amsmath}

\begin{document}
\renewcommand{\thefootnote}{\fnsymbol{footnote}}
\renewcommand{\d}{\textbf{\rm d}}

\begin{titlepage}

\begin{center}

\vspace{1cm}

{\Large {\bf Relic Density of Asymmetric Dark Matter with Breit-Wigner Enhancement  }}

\vspace{1cm}

{\bf Fangyu Liu,  Hoernisa Iminniyaz\footnote{Corresponding 
author, wrns@xju.edu.cn} Qiquan Li\footnote{Corresponding author, email:1015852575@qq.com}}


\vskip 0.15in
{\it
  {School of Physics Science and Technology, Xinjiang University,
  \\
Urumqi 830017, China}\\
}

\abstract{ The Breit-Wigner Enhancement mechanism provides a novel
  interpretation for the ``enhancement factor'' derived from data of the
  PAMELA, ATIC, and PPB-BETS experiments, and we further extends this research
  to the case of asymmetric Dark Matter. We investigate the impact of
  resonance mechanism to the asymmetric Dark Matter case and constrain the
  resonance model parameters to satisfy the ``enhancement factor" and
  the asymmetric dark matter model parameters to meet the upper
  bound requirement imposed by the astronomical experiment.}
 
\end{center}
\end{titlepage}
\setcounter{footnote}{0}

\section{Introduction}
The existence of Dark Matter (DM) has been confirmed through astronomical observations including galaxy rotation curves \cite{Pato:2015dua,Iocco:2015xga,Jiao:2023aci,Ou:2023adg}, cosmic microwave background radiation (CMB) \cite{Clowe:2006eq,Robertson:2016qef}, gravitational lensing effects, and large-scale structure formation \cite{Springel:2005nw}. Based on Planck 2018 observational data of CMB anisotropy, the relic density of the present cold DM is
\begin{equation}\label{eq:A1}
    \Omega_{\mathrm{CDM}} h^2 = 0.120 \pm 0.001 \quad (68\%\ \text{C.L.}),
\end{equation}
where $h = 0.674 \pm 0.005$ is the present Hubble expansion rate in units of
$100\ \mathrm{km\,s^{-1}\,Mpc^{-1}}$ \cite{Planck:2018vyg}. Despite such
compelling evidences for its existence, the fundamental nature of DM
remains one of the most profound unsolved mysteries in modern physics and
cosmology. Weakly Interacting Massive Particles (WIMPs) are usually considered
as good DM candidates, among them neutralino is the best one which is
Majorana particle for it the particle and anti-particle are the same.

Experimental searches for DM are of paramount importance, broadly categorized into the direct and indirect DM searches. Direct DM detection experiments observe elastic scattering events between DM particles and nuclei, while indirect searches aim to detect products from DM annihilation. In indirect detection experiments, the AMS-02 Collaboration's
2019 release of cosmic-ray positron flux measurements
demonstrated a significant spectral hardening phenomenon in the energy range
of approximately 25.2 GeV to 284 GeV \cite{AMS:2019rhg}. This feature
manifests as follows: at
energies above the characteristic cutoff of 25.2 GeV, the spectral index shows
a systematic decrease, resulting in a pronounced flattening of the
differential flux spectrum. This indicates enhanced relative abundance of
high-energy positrons within this energy interval compared to the low-energy
baseline. The experimental results, along with those of the PAMELA, ATIC, and
PPB-BETS experiments \cite{PAMELA:2008gwm,Chang:2008aa,PPB-BETS:2008zzu},
strongly imply the existence of a new positron source in cosmic rays. The
annihilation of DM can explain the above anomalies, however this
interpretation requires that the DM annihilation cross section  in the
galactic halo should be larger than the one which is predicted by the DM relic density measured by the WMAP experiment. Thus, there exists an enhancement factor. Currently, the
physical interpretations of the origin of the enhancement factor mainly
include two paths: one is the astrophysical mechanism, which amplifies the
signal through the spatial distribution effect of enhanced local DM density;
the other is the particle physics mechanism, with a typical representative
being the non-perturbative Sommerfeld enhancement 
effect caused by the long-range interactions between DM
particles \cite{Arkani-Hamed:2008hhe}.

For interpreting the enhancement factor, Ibe et al. proposed  the Breit-Wigner mechanism explanation\cite{Ibe:2008ye}. When the Resonance mass is just
below twice the DM mass, the annihilation cross section becomes
sensitive to the DM velocity and monotonically increases as the DM velocity decreases. They found that the cross section required from the dark matter density can be large enough to explain the PAMELA, ATIC, and
PPB-BETS results, without additional boost factor due to an overdense
region in the halo or the Sommerfeld enhancement\cite{Ibe:2008ye}. More studies on DM annihilation near the Resonance can be
found in
Refs. \cite{Guo:2009aj,Bi:2011qm,Duch:2017nbe,Ding:2021sbj,Belanger:2024bro,Cheng:2023dau}.

On the other hand, the fact that the DM density $\Omega_{\text{DM}}$ and baryonic matter density
   $\Omega_{\text{B}}$ are remarkably similar in order of magnitude
   $\Omega_{\text{DM}} \approx 4.7 \Omega_{\text{B}}$ proposes that DM likely
   possesses asymmetry akin to that of baryonic matter.
Refs.\cite{Shelton:2010ta,Hooper:2004dc,Abel:2006nv,Kaplan:2009ag} suggest
that DM may possesses asymmetry, similar to baryonic matter.
Asymmetric DM is also a possible candidate for DM.
Therefore, it is very important to extend the Breit-Wigner enhancement from
the case of symmetric DM to the case of asymmetric DM. We discuss the impact
of Breit-Wigner
enhancement on the relic density of asymmetric DM abundance and give the
constraints on the parameter spaces using the oberseved abundance of DM relic
density. 

The structure of this paper is organized as follows. In Sec.~\ref{sec:1}, we
first review the annihilation cross section for Breit-Wigner enhancement and
further derive the Boltzmann equations for asymmetric DM including this
mechanism. Then, we provide an analytical solution for the
relic density of asymmetric DM with Breit-Wigner enhancement. In Sec.~\ref{sec:2}, we constrain the parameter spaces using the current relic abundance of cold DM and at the same time, such constraints satisfy some indirect detection experiment requirements. Finally, the conclusion and summary are presented in Sec.~\ref{sec:3}.
\section{Relic abundance of asymmetric DM with Breit-Wigner enhancement}\label{sec:1}

In this subsection, we consider the Boltzmann equation for asymmetric DM with Breit-Wigner enhancement and present the evolution of DM density. 
The following Boltzmann equations are used to calculate the relic abundance of
asymmetric DM particles $\chi$ and their anti-particles $\bar{\chi}$,
\begin{equation} \label{eq:B1}
  \frac{\mathrm{d}n_{\chi}}{\mathrm{d}t}+3Hn_{\chi}=-\langle \sigma v \rangle
  (n_{\chi}n_{\bar{\chi}}-n_{\chi,\mathrm{eq}}n_{\bar{\chi},\mathrm{eq}})\,;
\end{equation}
\begin{equation} \label{eq:B2}
\frac{\mathrm{d}n_{\bar\chi}}{\mathrm{d}t}+3Hn_{\bar\chi}=-\langle \sigma v
  \rangle (n_{\chi}n_{\bar{\chi}}-n_{\chi,\mathrm{eq}}n_{\bar{\chi},\mathrm{eq}})\,,
\end{equation}
where the Hubble expansion rate $H = \pi T^{2}/M_{\rm
  Pl}\sqrt{g_{*}/90}$ with $M_{\rm Pl} = 2.4\times 10^{18}$ GeV is the reduced Planck mass, and $g_{*}$ is the effective number of relativistic degrees of freedom. $\langle \sigma v \rangle $ is the thermal average of
annihilation cross section times relative velocity of annihilating  
asymmetric DM particles and anti-particles. Here $n_{\chi}$ and $n_{\bar\chi}$
are the number densities of asymmetric DM particles and anti-particles. Their equilibrium values are given by:
$
n^{\rm eq}_{\chi} = g ~{\left( m T/2 \pi \right)}^{3/2} {\rm exp}[{(-m + \mu_{\chi})/T}]\,,\,\,\,\,\,
n^{\rm eq}_{\bar\chi} =  g ~{\left( m T/2 \pi \right)}^{3/2} {\rm exp}[{(-m - \mu_{\bar\chi})/T}]\,, $
where $m$ is the mass of the asymmetric DM, $\mu_{\chi}$ and $\mu_{\bar{\chi}}$ are the chemical potentials of the particles and anti-particles, respectively, with $\mu_{\bar{\chi}} = -\mu_{\chi}$ in equilibrium, and $g$ is the number of internal degrees of freedom of the particles.

Using the dimensionless variable $Y = n/s$ and $x$ with $s =
(2\pi^2/45)g_{*s}T^3$ being the entropy density and $g_{*s}$ the effective
entropy degrees of freedom, and again subtracting Boltzmann equations for
$Y_{\chi}$ and $Y_{\bar\chi}$, we can rewrite the Boltzmann
Eqs.(\ref{eq:B1}),(\ref{eq:B2})
as 
\begin{equation} \label{eq:B3}
\frac{{{\rm d}Y_{\chi}}}{{\rm d}x}=-\frac{\lambda\langle \sigma v\rangle}{x^2}
      \left(Y_{\chi}^{2}-\eta Y_{\chi}-P\right)\,;
    \end{equation}
\begin{equation} \label{eq:B4}
\frac{{\rm d}Y_{\bar{\chi}}}{{\rm d}x}=-\frac{\lambda\langle \sigma v\rangle}{x^2}
      \left(Y_{\bar{\chi}}^{2}+\eta Y_{\bar{\chi}}-P\right),
\end{equation}
here $\eta \equiv Y_{\chi}-Y_{\bar{\chi}}$  is a constant, $\lambda = 1.32\, m
M_{\mathrm{Pl}}\sqrt{g_{*}}$ ,
$P=Y_{\chi,\text{eq}}Y_{\bar{\chi},\text{eq}}=(0.145g_{\chi}/g_{*})^{2}x^{3}e^{-2x}$
with $Y_{\bar{\chi},eq}=-\eta/2+\sqrt{\eta^{2}/4+P}$. We
noted that $P$ doesn't depend on the chemical potential $ \mu_{\chi} $.

We use the scattering cross section formula of scalar Resonance, given by Masahiro Ibe et al.\cite{Ibe:2008ye}.
For the scalar Resonance, two auxiliary parameters $\delta$ and $\gamma$ are
introduced to give the scattering cross section of the narrow Resonance $R$,
\begin{eqnarray} \label{eq:B5}
\delta = 1 - M^{2}/4m_{\chi}^{2} \quad \text{and} \quad \gamma = \Gamma/M,
\end{eqnarray}
where $M$ and $\Gamma$ represent the mass and the decay width of the Resonance, respectively. The annihilation cross section is given by
\begin{eqnarray} \label{eq:B6}
\sigma = \frac{16\pi}{M^{2}\bar{\beta}_{i}\beta_{i}} \frac{\gamma^{2}}{(\delta + v^{2}/4)^{2} + \gamma^{2}} B_{i} B_{f},
\end{eqnarray}
where $ B_{i} $ and $ B_{f} $ are the branching fractions of the Resonance
into the initial and final channels, respectively. $\bar{\beta}_{i} = \sqrt{1
  - 4m^{2}/M^{2}}$ and $\beta_{i} = \sqrt{1 - 4m^{2}/E_{\mathrm{cm}}^{2}}$ are
the initial state phase space factors evaluated on Resonance and the phase
space factors at the center of mass energy of the collision, respectively.
Even though an un-physical pole exists for $\delta > 0$ when $2m > M$, the ratio $B_i/\bar{\beta}_i$ defined via analytic continuation from the physical region $\delta< 0$ when $M<2m$ and the corresponding cross section remains well-defined in both regions.
For non-relativistic velocities, the thermal average of the annihilation cross
section multiplied by the relative velocity of asymmetric DM can be well approximated by Gaussian integration as:
\begin{eqnarray} \label{eq:B7}
\langle\sigma v \rangle \simeq \frac{32\pi}{M^{2}\bar{\beta}_{i}} \frac{\gamma^{2}}{(\delta + \zeta\,x^{-1})^{2} + \gamma^{2}} B_{i}B_{f},
\end{eqnarray}
where a parameter $\zeta \approx 1/\sqrt{2}$ provides the optimal fit to numerical results for $v_{0} \ll 1$ and $\delta > 0$, with $x =v_{0}^{-2}=m/T$ being the mass to temperature ratio. From Eq.(\ref{eq:B7}), we can find that due to the presence of the denominator term $x^{-1}$, the cross section exhibits an enhancement effect at lower temperatures when $\delta$ and $\gamma$ are small.
After inserting Eq.(\ref{eq:B7}) into
Boltzmann Eqs.(\ref{eq:B3}),(\ref{eq:B4}), then 

\begin{equation} \label{eq:B8}
\frac{{\rm d}Y_{\chi}}{{\rm d}x}=-\frac{\lambda\sigma_0}{x^2}
      \frac{\left(\delta^{2} + \gamma^{2}\right)}{\left(\delta + \zeta x^{-1}\right)^{2} + \gamma^{2}} \left(Y_{\chi}^{2}-\eta Y_{\chi}-P\right)\,;
\end{equation}
\begin{equation} \label{eq:B9}
\frac{{\rm d}Y_{\bar{\chi}}}{{\rm d}x}=-\frac{\lambda\sigma_0}{x^2}
      \frac{\left(\delta^{2} + \gamma^{2}\right)}{\left(\delta + \zeta x^{-1}\right)^{2} + \gamma^{2}} \left(Y_{\bar{\chi}}^{2}+\eta Y_{\bar{\chi}}-P\right)\,,
\end{equation}
here, $\sigma_{0} = \langle\sigma v \rangle_{T=0} = 32\pi B_{i} B_{f}\gamma^{2}  / [M^{2} \bar{\beta}_{i}(\delta^{2} + \gamma^{2})]$.

\begin{figure}[h]
  \begin{center}
    \hspace*{-0.5cm} \includegraphics*[width=8cm]{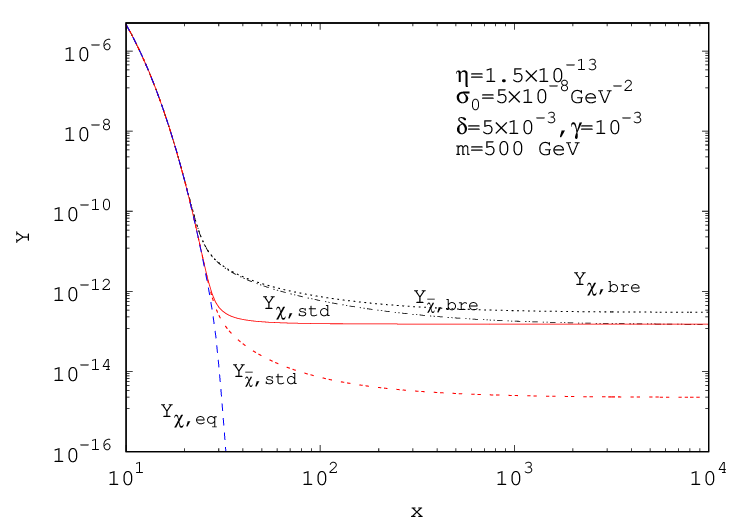}
         \caption{\label{fig:yxadmbr} \footnotesize
       $Y$ as a function of $x$ with the Breight-Wigner enhancement.  Here,
      $g=2$, $g_*=90$.              }
   \end{center}
\end{figure}

For the standard cosmological model, asymmetric DM particles were initially in thermal equilibrium in the early universe. As the universe expanded and the temperature continuously decreased, these particles gradually deviate from thermal equilibrium when the annihilation
rate fell below the Hubble expansion rate. The asymmetric DM particles then entered a
freeze-out state, rapidly forming a nearly constant relic, and the freeze-out
time $x_f$ can typically be defined by the equation $Y_{\bar{\chi}} -
Y_{\bar{\chi},\mathrm{eq}}=\xi Y_{\bar{\chi},\text{eq}}(x_f)$ where
$\xi=\sqrt{2}-1$ in the standard model. However, in asymmetric DM with Breit-Wigner
enhancement, the situation is different, as shown in Fig.\ref{fig:yxadmbr}. In
this picture, we plot the asymmetric DM abundance $Y$ as a function of $x$ in two different
scattering scenarios, Resonance scattering and S-wave scattering.
Here, S-wave
cross section $\sigma_s$ is taken as $\sigma_s=\sigma_0$ 
which is same with the Resonance case. Taking this choice, the thermal-equilibrium
decoupling point with Resonance could be earlier than the S-wave case,
since
the former interaction rate $n_{\bar{\chi},\rm eq}\langle \sigma v \rangle$ is
smaller than the latter when making a comparison of the interaction rate of
DM particle with the expansion rate of the universe $\sqrt{8\pi
  G\rho_{\rm radiation}/3}$. Using this criterion, the thermal equilibrium decoupling point can be expressed as 
\begin{equation}\label{eq:B10}
\sqrt{\frac{x_{f0}}{{x}_{f_R}}}\exp({x}_{f_R}-x_{f0})=\frac{\left(\delta^{2} + \gamma^{2}\right)}{\left(\delta + \zeta {x}_{f_R}^{-1}\right)^{2} + \gamma^{2}}
\end{equation}  
where ${x}_{f_R}$ and $x_{f0}$ is the thermal decoupling point of Resonance case and S-wave case respectively.
Since the cross section depends on the temperature and increases when the temperature decreases, the Hubble expansion rate can not completely dominate over the interaction rate throughout the freeze-out period. Thus, after thermal-decoupling, the freeze-out of the relic lasts for a much longer period. Of course, with this parameter choice ($\sigma_s=\sigma_0\leqslant\sigma$), the relic density of Resonance case is higher than the S-wave case. This in turn implies that, under the constraint by the astrophysical observational relic density, the larger current cross section $\sigma_0$ is allowed in Resonance mechanism, when comparing to the S-wave case.

We can use the method as in \cite{Iminniyaz:2013cla} to get an analytical
solution for the relic density of asymmetric DM with Breit-Wigner enhancement. By using $\Delta_{\bar{\chi}} = Y_{\bar{\chi}} - Y_{\bar{\chi},\mathrm{eq}}$, we simplify the Boltzmann Eq.(\ref{eq:B9}) to
\begin{eqnarray} \label{eq:C1}
\frac{{\rm d}\Delta_{\bar{\chi}}}{{\rm d}x} = -\frac{{\rm
  d}Y_{\bar{\chi},\mathrm{eq}}}{{\rm d}x} -\frac{\lambda\,\sigma_0}{x^2}
      \frac{\left(\delta^{2} + \gamma^{2}\right)}{\left(\delta + \zeta x^{-1}\right)^{2} + \gamma^{2}} \left[ \Delta_{\bar{\chi}}(\Delta_{\bar{\chi}} + 2Y_{\bar{\chi},\mathrm{eq}}) + \eta\Delta_{\bar{\chi}} \right].
\end{eqnarray}
For high temperature, the above equation can be simplified to obtain
\begin{eqnarray} \label{eq:C2}
\Delta_{\bar{\chi}}\simeq\frac{2x^{2}P}{\lambda \,\sigma_0}\frac{\left(\delta + \zeta x^{-1}\right)^{2} + \gamma^{2}}{\left(\delta^{2} + \gamma^{2}\right)\left(\eta^{2}+4P\right)}.
\end{eqnarray}
From Eq.(\ref{eq:C2}) and $Y_{\bar{\chi}} - Y_{\bar{\chi},\mathrm{eq}}=\xi
Y_{\bar{\chi},\text{eq}}(\widetilde{x}_f)$, we can determine that in
asymmetric DM with Breit-Wigner enhancement, the freeze-out process ends at
temperature $ \widetilde{x}_f $.

For sufficiently low temperature, $ x \gg \widetilde{x}_f $, Eq. (\ref{eq:C1}) becomes
\begin{eqnarray} \label{eq:C3}
\frac{{\rm d}\Delta_{\bar{\chi}}}{{\rm d}x}=-\frac{\lambda\, \sigma_0}{x^2}
      \frac{\left(\delta^{2} + \gamma^{2}\right)}{\left(\delta + \zeta x^{-1}\right)^{2} + \gamma^{2}}\left(\Delta_{\bar{\chi}}^{2}+\eta\Delta_{\bar{\chi}}\right),
\end{eqnarray}
By integrating above equation from $\widetilde{x}_f$ to $\infty$, we
obtain the final abundance for asymmetric DM  anti-particle as
\begin{eqnarray} \label{eq:C4}
Y_{\bar{\chi}}(x\rightarrow\infty)=\frac{\eta}{\exp\left[1.32\,\eta\,m_{\chi}M_{\mathrm{Pl}}\,\sqrt{g_*}\,I(\widetilde{x}_f)\right]-1}\,,
\end{eqnarray}
where
\begin{eqnarray} \label{eq:C5}
I(\widetilde{x}_f) =  \int\limits_{\widetilde{x}_f}^{\infty}\frac{\left(\delta^{2} + \gamma^{2}\right)}{\left(\left(\delta + \zeta x^{-1}\right)^{2} + \gamma^{2}\right)x^2}\,dx
 = \frac{\left(\delta^{2}+\gamma^{2}\right)}{\zeta \gamma} \arctan \left(\frac{\zeta \gamma}{\left(\delta^{2}+\gamma^{2}\right) \widetilde{x}_f+\delta \zeta}\right).
\end{eqnarray}
Then the relic abundance for $\chi$ particles is given by
\begin{eqnarray} \label{eq:C6}
Y_{\chi}(x\rightarrow\infty)=\frac{\eta}{1-\exp\left[-1.32\,\eta\,m_{\chi}\,M_{\mathrm{Pl}}\,\sqrt{g_*}\,I(\widetilde{x}_f)\right]}\,.
\end{eqnarray}

The predicted relic abundance for asymmetric DM is expressed as
\begin{eqnarray} \label{eq:C7}
\Omega_{\mathrm{DM}}h^{2} & = & \frac{m_{\chi}s_{0}\left[Y_{\chi}\,(x\rightarrow\infty)+Y_{\bar{\chi}}\,(x\rightarrow\infty)\right]h^{2}}{\rho_{\mathrm{crit}}}\,.
\end{eqnarray}
where $s_{0}=2.9\times 10^{3}~\mathrm{cm}^{-3}$ is the current entropy density and $\rho_{\mathrm{crit}}=3M_{\mathrm{Pl}}^{2}H_{0}^{2}$ is the critical density.

\section{Constraints on parameter spaces}\label{sec:2}

In this section we use the observed DM relic density provided by Planck data
to find constraints on the parameter spaces.
We first obtain constraints on the Resonance model parameters in order to
satisfy the astronomical experimental requirement that the current cross
section of
DM in the galactic halo should be much larger than the one predicted by the
observed DM
relic density. Then, we obtain  constraints on the current cross section $\sigma_0$ to meet  the upper bound requirement imposed by the Fermi-LAT and CALET experiments.

For late time, the
Boltzmann equation (\ref{eq:B4}) for asymmetric DM S-wave case becomes
\begin{equation}\label{eq:D1}
\frac{{\rm d}Y_{\bar{\chi}}}{{\rm d}x}=-\frac{\lambda}{x^2} \sigma_s(Y_{\bar{\chi}}^2+\eta Y_{\bar{\chi}})\,.
\end{equation}
where $\sigma_s$ is constant, $P$ in Eq.(\ref{eq:B4}) is neglected as a second-order small quantity. Integrating Eq.(\ref{eq:D1}) yields 
\begin{equation}
{\rm ln}\frac{Y_{\bar{\chi}}(\infty)}{Y_{\bar{\chi}}(\infty)+\eta}=-\eta\lambda\sigma_s\frac{1}{x_f}\,.
\end{equation}
The above equation gives the expression 
 \begin{equation}\label{lambdasigma}
\lambda\sigma_s=\frac{x_f}{\eta}{\rm ln}[1+\frac{2\eta k}{\Omega h^2-\eta k}]\,,
\end{equation} 
where $k=m_\chi s_0 h^2/\rho_{\rm crit}$. In quantitatively analyzing  the effect
of Resonance on the current-time annihilation cross section in asymmetric DM
model, the above expression can be used. Defining  the ratio of the
annihilation cross sections in the
Resonance and S-wave case as the enhancement factor
$\beta\equiv\sigma_0/\sigma_s$. Here the cross section $\sigma_0$ of Resonance
case, and the cross section  $\sigma_s$ of S-wave case, are required to,
respectively and simultaneously produce the experimental observational  DM
relic abundance. This requirement and the physical property of the origin of expression Eq.\eqref{lambdasigma} support the validity that expression Eq.(\ref{lambdasigma}) can be utilized in the following calculation.

Thus, the Boltzmann equations (\ref{eq:B8}),(\ref{eq:B9}) with Resonance  can
be expressed as
\begin{equation}\label{omegaY}
\frac{{\rm d}Y_{{\chi}}}{{\rm d }x}=-\frac{\beta}{x^2}  \frac{x_f}{\eta}{\rm ln}[1+\frac{2\eta k}{0.120-\eta k}]
      \frac{\left(\delta^{2} + \gamma^{2}\right)}{\left(\delta + \zeta
          x^{-1}\right)^{2} + \gamma^{2}} \left(Y_{{\chi}}^{2}-\eta
        Y_{{\chi}}-P\right)\,;
\end{equation}
\begin{equation}\label{omegaYbar}
 \frac{{\rm d }Y_{\bar{\chi}}}{{\rm d}x}=-\frac{\beta}{x^2}  \frac{x_f}{\eta}{\rm ln}[1+\frac{2\eta k}{0.120-\eta k}]
      \frac{\left(\delta^{2} + \gamma^{2}\right)}{\left(\delta + \zeta x^{-1}\right)^{2} + \gamma^{2}} \left(Y_{\bar{\chi}}^{2}+\eta Y_{\bar{\chi}}-P\right)\,.
\end{equation}
    
\begin{figure}[h]
  \begin{center}
    \hspace*{-0.5cm} \includegraphics*[width=8.7cm]{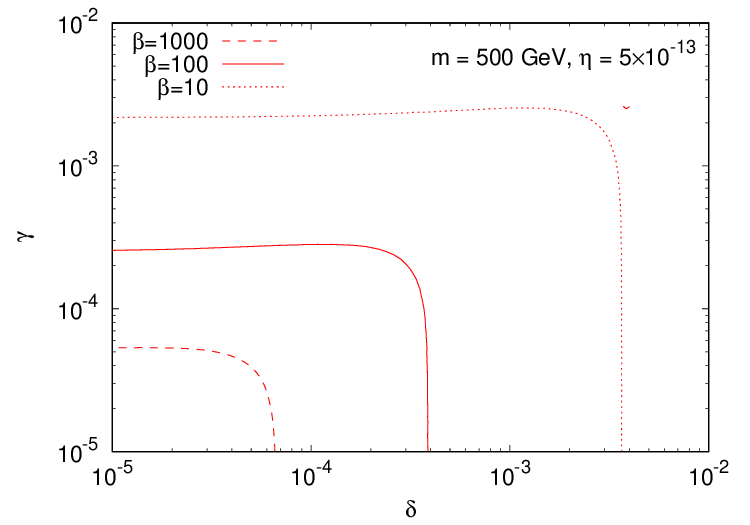}
    \caption{\label{fig:cross} \footnotesize
      The boost factor on the panel $(\delta,\ \gamma)$.
      Here, $\Omega_{\rm DM} h^2 = 0.120$,
      $g = 2$, $g_* = 90$, $m=500 \rm GeV$, $x_f=22$.}
     \end{center}
\end{figure}

When the above equation yields the observed DM relic density data, the
parameter $\beta$ represents the annihilation cross section
boost factor in asymmetric DM.  Utilizing the aforementioned equations Eq.\eqref{omegaY} and
Eq.\eqref{omegaYbar}, we plot the constraints on parameters of asymmetric DM
with Resonance mechanism, required to satisfy the 
observational relic density of DM and the annihilation strength in the galactic
halo at the same time. Fig.(\ref{fig:cross}) plots the constraints on
parameter spaces. The boost factor is large when $\delta$ and $\gamma$ are small.

Although the relic abundance from a cosmoscopic perspective has frozen out in
the early universe, the annihilation process is still occurring in local high
DM density region, which can hypothetically be utilized to explain the origin
of the certain Standard Model Particle as positon excess signal. Mutually, the
observational data places constraints on the parameter spaces of the specific DM model. Fermi Large Area Telescope gives the upper limits on the annihilation  cross section of a symmetric DM candidate through the gamma
ray observations of Milky Way dwarf spheroidal satellite galaxies \cite{Fermi-LAT:2015att}.
Ref.\cite{Hiroshima:2025jyz} computes the upper limit on the cross section of self-conjugate DM which explain the observed electron and positron fluxes from CALET and other experiments. In this subsection, we apply these results to explore the asymmetric DM with Resonance. For the symmetric DM relic, the present DM annihilation rate is 
$\Gamma_{\rm sym}=\langle\sigma_{\rm self}v\rangle(n^2_{\rm
  DM})/2=\langle\sigma_{\rm self}v\rangle(\rho^2_{\rm DM}/2m^2)$, here
$\sigma_{\rm self}$ is the annihilation cross section for symmetric DM. For
asymmetric DM models, 
$\Gamma_{\rm asym}=\langle\sigma_{\chi\bar{\chi}}v\rangle
n_{\chi}n_{\bar{\chi}}=\langle\sigma_{\chi\bar{\chi}}v\rangle(\rho_{\chi}\rho_{\bar{\chi}}/m^2)=\langle\sigma_{\chi\bar{\chi}}v\rangle
(\rho^2_{\rm DM}/m^2)  Y_{\chi}Y_{\bar{\chi}}/(Y_{\chi}+Y_{\bar{\chi}})^2$,
where $\rho_{\rm DM}=\rho_{\chi}+\rho_{\bar{\chi}}$. Since the present DM relic annihilation rate is physical quantity, a comparison can be made between the symmetric and asymmetric case. Then, the ratio of the annihilation rate for asymmetric DM and symmetric DM with upper bound obtained from the observation data should be less than 1 \cite{Gelmini:2013awa},
\begin{equation}\label{F-ratio1}
\frac{\Gamma_{\rm asym}}{\Gamma_{\rm
    up}}=\frac{\langle\sigma_{\chi\bar{\chi}}v\rangle}{\langle\sigma
  v\rangle_{\rm up}}\frac{2Y_{\chi}Y_{\bar{\chi}}}{(Y_{\chi}+Y_{\bar{\chi}})^2}<1\,,
\end{equation}
The aforementioned expression can be rewritten as 
\begin{equation}\label{F-ratio2}
\frac{\Omega_{\rm DM}h^2}{2.76\times 10^8m_{\chi}}\left(1-2\frac{\langle\sigma
  v\rangle_{\rm Fermi}}{\langle\sigma_{\chi\bar{\chi}}
  v\rangle}\right)^{1/2}<\eta \,.
\end{equation} 
%
For the computation of constraints on $\eta$ with $\sigma_0$, we use the similar method as above. In symmetric S-wave case, we can obtain
\begin{equation}
\lambda \sigma_{\rm sym}=\frac{1.32\times 8.5\times 10^{-11}m  M_{\rm Pl} x_f^\prime}{\Omega h^2}\,,
\end{equation}
here $x_f^\prime$ is the freeze-out temperature for symmetric DM. Thus, the Boltzmann Eq.(\ref{omegaY}) can be written as follows,
\begin{equation}
\frac{{\rm d}Y_{{\chi}}}{{\rm d}x}=-\frac{\beta^\prime}{x^2}  \frac{1.32\times
  8.5\times 10^{-11}m  M_{\rm Pl} x_f^\prime}{\Omega h^2}
      \frac{\left(\delta^{2} + \gamma^{2}\right)}{\left(\delta + \zeta x^{-1}\right)^{2} + \gamma^{2}} \left(Y_{{\chi}}^{2}-\eta Y_{{\chi}}-P\right)\,,
\end{equation}
where $\beta^\prime\equiv \sigma_0/\sigma_{\rm sym}$. The parameter
$\sigma_{\rm sym}$ is the symmetric S-wave cross section and is chosen to be
the canonical value of $3\times 10^{-26}\ \rm cm^3 s^{-1}$ which is
independent of mass $m$.
\begin{figure}[h]
  \begin{center}
    \hspace*{-0.5cm} \includegraphics*[width=8cm]{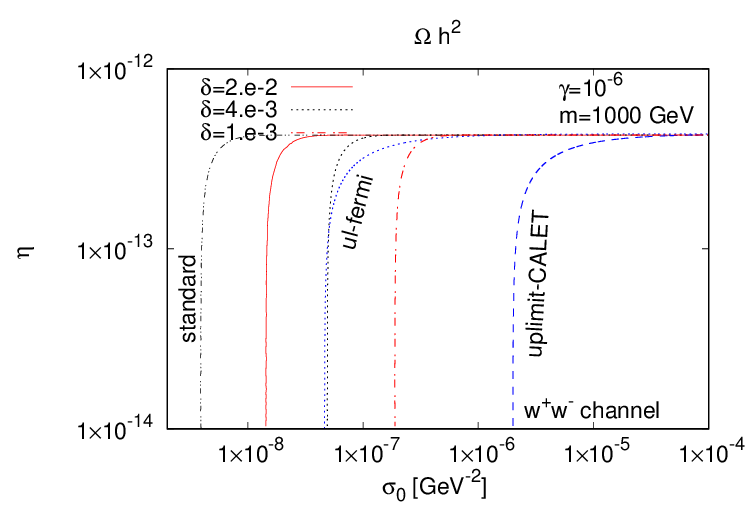}
     \caption{\label{fig:omegamc} \footnotesize
      Contour plots of  $\eta$ and $\sigma_0$ ; when 
    $\Omega_{\rm DM} h^2 = 0.120$ \cite{Planck:2018vyg}.
      Here $m = 1000$ GeV, $g = 2$, $g_* = 90$, $x_f^\prime=22$.         }
   \end{center}
 \end{figure}
 
Fig.(\ref{fig:omegamc}) plots the constrains on parameter spaces spanned by
$\sigma_0$ and $\eta$. The region to the left of the lines labelled by $ul({\rm
uplimit})$ is
the allowed parameter region, as the region to the right of those lines yields
the higher DM annihilation rate. Here, we use the conclusion that $\sigma_{\rm
  up}= 2.6867\times 10^{-25}\rm\ cm^3s^{-1}$ for six years Fermi-LAT Milky Way
Dwarf Spheroidal Galaxies data and $\sigma_{\rm up}= 1.16\times 10^{-23}\rm\
cm^3s^{-1}$ for CALET electron flux data in $W^+W^-$ channel
\cite{Fermi-LAT:2015att,Hiroshima:2025jyz}.
The line denoted by ``standard" represents the asymmetric DM  S-wave case. From
Fig.(\ref{fig:omegamc}), we can see that compared to the asymmetric DM S-wave
case, the Resonance case can approach the experimental limit more closely,
which means that the broader detection range is allowed. For example, with $\eta$
fixed at $10^{-13}$ and the other parameters as presented, the asymmetric DM
S-wave case gives the cross section $3.96\times 10^{-9}\ \rm GeV^{-2}$, while
the asymmetric DM with Resonance gives the larger cross section $1.46\times 10^{-8}\ \rm GeV^{-2}$ for $\delta=2\times 10^{-2}$, $4.98\times 10^{-8}\ \rm GeV^{-2}$ for $\delta=4\times 10^{-3}$ and $1.92\times 10^{-7}\ \rm GeV^{-2}$ for $\delta=1\times 10^{-3}$. 

\section{Summary and conclusions}\label{sec:3}
In this paper, we extend the study of the evolution of symmetric DM
with Breit-Wigner enhancement to the asymmetric DM scenarios. The analytical
solutions for relic density in such case is obtained and density evolution is presented.
Since the annihilation in the early universe is suppressed by the Breit-Wigner temperature-dependent term, the annihilation cross section required by observed DM relic density should be larger, which can be consistent with the excess of positron experiments results. The constraints on parameters of Resonance model is obtained in 
order to satisfy such consistency i.e. simultaneously fulfill the annihilation
strength in the galactic halo
from astronomical observation and the relic density from Planck observation. Then, we apply the upper limit of the cross section from Fermi-Lat and CALET data to
constrain the asymmetric DM model parameters, the current cross section vs asymmetry. Meeting the experimental upper bound requirement, Resonance mechanism can approach the limit more closely than the S-wave case.

\section*{Acknowledgments}

The work is supported by the National Natural Science Foundation of China (Grant No. 12463001).

\end{document}